\documentclass[12pt]{iopart} 
\usepackage{iopams}

\usepackage{bm}
\usepackage{graphicx} 
\usepackage{hyperref}
\usepackage{mathrsfs}
\usepackage{multirow}    
\usepackage{dsfont}    
\usepackage{bbm}

\newcommand{\ud}{\mathrm{d}}

\newcommand{\calO}{\mathcal{O}}

\newcommand{\text}[1]{{\rm{#1}}}

\newcommand{\eqref}[1]{(\ref{#1})}

\usepackage{ulem}
\usepackage[usenames]{color}

\begin{document}

\title[Sun/Moon gravitational redshift tests]{Analysis of Sun/Moon
  Gravitational Redshift tests with the STE-QUEST Space Mission}

\author{Peter Wolf\,$^1$ and Luc Blanchet$^2$}

\address{$^1$ LNE-SYRTE, Observatoire de Paris, PSL Research
  University, CNRS, Sorbonne Universit\'es, UPMC Univ. Paris 06, 61
  avenue de l'Observatoire, 75014 Paris, France}

\address{$^2$ GRECO, Institut d'Astrophysique de Paris, CNRS UMR 7095,
  Universit\'e Pierre \& Marie Curie, 98$^{\rm bis}$ boulevard Arago,
  75014 Paris, France} 

\ead{peter.wolf@obspm.fr, blanchet@iap.fr}
\vspace{10pt}
\begin{indented}
\item[]\today
\end{indented}

\begin{abstract}
The STE-QUEST space mission will perform tests of the gravitational
redshift in the field of the Sun and the Moon to high precision by
frequency comparisons of clocks attached to the ground and separated
by intercontinental distances. In the absence of Einstein equivalence
principle (EP) violation, the redshift is zero up to small tidal
corrections as the Earth is freely falling in the field of the Sun and
Moon. Such tests are thus null tests, allowing to bound possible
violations of the EP. Here we analyze the Sun/Moon redshift tests
using a generic EP violating theoretical framework, with clocks
minimally modelled as two-level atoms. We present a complete
derivation of the redshift (including both GR and non-GR terms) in a
realistic experiment such as the one envisaged for STE-QUEST. We point
out and correct an error in previous formalisms linked to the atom's
recoil not being properly taken into account.
\end{abstract}

%
%
%
%
%

\section{Introduction} 
\label{sect:intro}

It is well known that two clocks fixed on the Earth's surface, when
compared to each other, do not display a frequency difference due to
external masses (Sun, Moon, Planets) at first order in $\Delta
U_\text{ext}/c^2$ --- a fact sometimes referred to as the
``\textit{absence of the Noon-Midnight
  redshift}''~\cite{H61,AW13}. Here $U_\text{ext}$ is the Newtonian
potential of external masses, and $c$ is the speed of light in
vacuum. This is of course a direct consequence of the equivalence
principle of General Relativity (GR), as the Earth is freely falling
in the field of external masses, and there is no effect due to the
rotation of the Earth in a local inertial frame. Thus a co-moving
frame is inertial and physics is described in that frame by the laws
of Special Relativity.

Instead, only tidal terms can be observed, which are a factor $\sim
r_\text{E}/R_\oplus$ smaller, where $r_\text{E}$ is the Earth radius,
and $R_\oplus$ is the distance between the Earth centre and the
barycentre of the external masses. Hence in the case of the Sun the
redshift scales like $\sim G M_\odot r^2_\text{E}/(R^3_\oplus c^2)$
and is very small indeed. For clocks on the Earth's surface such tidal
terms do not exceed a few parts in $10^{17}$ in fractional frequency,
and are thus barely measurable with today's best clocks, whilst first
order terms due to the Sun, for example, would be about 5 orders of
magnitude larger.

In 2010 the ``Space-Time Explorer and QUantum Equivalence principle
Space Test'' (STE-QUEST) project~\cite{STEQUEST} was proposed to the
European Space Agency,\footnote{STE-QUEST is currently competing for
  the ESA medium size mission call M5.} which included a test of the
gravitational redshift in the field of the Sun/Moon at first order in
$\Delta U_\text{ext}/c^2$. The idea is to compare, \textit{via} the
STE-QUEST satellite, two clocks attached to the Earth and separated by
intercontinental distances. The comparison is made using microwave
links in common-view mode. The test boils down to a search for a
periodic signal with known frequency and phase in the clock comparison
data. Roughly speaking the proposed experiment would test the
gravitational redshift from external masses at a level of $\sim
(\Delta f/f)(c^2/\Delta U_\text{ext})$, where $\Delta f/f$ is the
uncertainty in fractional frequency of the used clocks and frequency
transfer techniques.

During the assessment study of STE-QUEST~\cite{PRC}, and on several
subsequent occasions, it was argued that such a test was impossible
because of the absence of Noon-Midnight redshift --- there is no
Sun/Moon redshift to measure at first order in $\Delta
U_\text{ext}/c^2$, instead only tidal terms can be measured, and
therefore the proposed test would perform less well, roughly by a
factor $\sim r_\text{E}/R_\oplus$.

In this paper we show that such claims are unfounded. Indeed, the
conclusion is only valid when the equivalence principle (EP) is
satisfied. As tests of the gravitational redshift are precisely
testing that hypothesis, \textit{i.e.} consider a possible violation
of the EP, the absence of the Noon-Midnight redshift cannot be taken
for granted in that situation. More generally, such tests search for a
general anomalous coupling between the clock internal energy and the
source of the gravitational field (Sun, Moon, ...), couplings which
would lead precisely to slight deviations from the null-redshift due
to external masses. We show within a simple but general
theoretical framework~\cite{Dicke,Nordtvedt,Haugan,Will} that such
deviations are proportional directly to $\Delta U_\text{ext}/c^2$, and
not only to tidal terms. At this occasion we point out and correct an
error in the previous formalism of Ref.~\cite{Haugan} (see
also~\cite{Will}), in which the atom's recoil velocity was not
properly taken into account, resulting in a wrong sign for the
second-order Doppler effect in the redshift formula.

More generally, we believe that our results will be of interest to
anyone interested in experimental tests of the equivalence principle,
as they present a simple and straightforward way of modelling such
experiments and evaluating their respective merits and underlying
theoretical connections. We also take this opportunity to present a
general and straightforward derivation of the standard GR result.

For simplicity our derivations will be restricted to a rather
  elementary (quasi-Newtonian, with almost no quantum mechanics
  involved) Lagrangian-based formalism, which is directly issued from
  Refs.~\cite{Dicke,Nordtvedt,Haugan,Will}. However we expect that our
  conclusions will remain unchanged when working within the very broad
  and sophisticated Standard Model Extension (SME)
  formalism~\cite{KS89,BK06,KT11}, as well as other EEP violating
  frameworks such as~\cite{Dequiv,DDono10}. A full analysis of STE-QUEST in those frameworks is beyond the scope of this paper, but will be subject of future work.

Our article is organized as follows: We first briefly describe in
Sec.~\ref{sect:theory} the EP violating theoretical framework to be
used. Then we apply it in Sec.~\ref{sect:simple} to a highly
simplified ``Gedanken'' experiment that allows us to derive our main
conclusion in a few simple steps. This section can be skipped by readers that are interested only in a fully realistic and complete scenario. Such a scenario is treated in Sec.~\ref{sect:full} with a complete calculation for a realistic
experiment as envisaged, for example, in the STE-QUEST project,
including all relevant terms, both for the standard GR contribution
and for the EP-violating terms. We finish with some concluding remarks
in Sec.~\ref{sect:conclusion}.

\section{Theoretical framework}
\label{sect:theory}

In this paper we shall perform our analysis using a broad class of
equivalence principle (EP) violating frameworks, encompassing a large
class of non-metric gravity theories and consistent with Schiff's
conjecture~\cite{Schiff}. This class of theories is defined by a
modified Lagrangian describing some physical composite system
(\textit{e.g.} an atom or an atomic clock), in which the coupling
between gravitation and different types of internal mass-energies is
generically not universal, \textit{i.e.} depends on the system and the
type of energy in question. Assimilating the system to a point mass in
the gravitational field, and working in the non relativistic
approximation, we have the Lagrangian
\begin{equation}\label{L0} L = - m c^2 + \frac{1}{2} m \,\bm{V}^2 
+ m \,U(\bm{X})\,,
\end{equation}
where $m$ is the total mass of the system, $U\equiv U_\text{ext}$ is
the Newtonian gravitational potential at the position $\bm{X}=(X^i)$
in a global frame, \textit{e.g.} centred on the Solar System
barycenter, and $\bm{V}=(V^i)$ is the velocity of the system in that
frame. Following the ``Modified Lagrangian framework'' (slightly
adapted from Refs.~\cite{Dicke,Nordtvedt,Haugan,Will}) the violation
of the EP is encoded into an abnormal dependence of the mass of the
system on the velocity $\bm{V}$ and position $\bm{X}$. This implies a
violation, respectively, of the local Lorentz invariance (LLI) and
local position invariance (LPI) aspects of the Einstein equivalence
principle~\cite{Dicke,Nordtvedt,Haugan,Will}. Such abnormal dependence
is due to a particular internal energy $E_\text{X}$ in the system,
hence
\begin{equation}\label{m} 
m(\bm{X}, \bm{V}) = \overline{m} + \frac{1}{c^2}\Biggl[
  E_\text{X}(\bm{X}, \bm{V}) + \sum_{\text{Y}\not=
    \text{X}}\overline{E}_\text{Y}\Biggr]\,,\end{equation}
while the other types of energies $\overline{E}_\text{Y}$ behave
normally. Here $\overline{m}$ is the sum of the rest masses of the
particles constituting the system. We shall pose
\begin{equation}\label{EX} 
E_\text{X}(\bm{X}, \bm{V}) = \overline{E}_\text{X} - \frac{1}{2}\delta
m_\text{I}^{ij}\,V^i V^j - \delta
m_\text{P}^{ij}\,U^{ij}(\bm{X})\,,\end{equation}
where $U^{ij}$ denotes the usual Newtonian tensor
potential,\footnote{Thus, $U^{ij}(\bm{X})\equiv G\int
  \ud^3x\,\rho(\bm{x})
  \frac{(X-x)^i(X-x)^j}{\vert\bm{X}-\bm{x}\vert^3}$, whose trace is
  $U^{ii}=U$.} and $\delta m_\text{I}^{ij}$ and $\delta
m_\text{P}^{ij}$ are two constant tensors describing the EP
violation. The sum of all ordinary energies in the system is then
$\overline{E}=\overline{E}_\text{X}+\sum_{\text{Y}\not=
  \text{X}}\overline{E}_\text{Y}$, and we may define
$m_0=\overline{m}+\overline{E}/c^2$. The Lagrangian~\eqref{L0} now
becomes
\begin{equation}\label{L} L = - m_0 c^2 + \frac{1}{2} m_0 
\left(\delta^{ij} + \beta_\text{I}^{ij}\right) V^i V^j +
m_0\left(\delta^{ij} + \beta_\text{P}^{ij}\right)
\,U^{ij}(\bm{X})\,,\end{equation}
where we have neglected small higher-order relativistic
corrections. For later convenience we defined two LLI and LPI
violating parameters $\beta_\text{I}^{ij}$ and $\beta_\text{P}^{ij}$
by
\begin{equation}\label{beta} \beta_\text{I}^{ij}=\frac{\delta
  m_\text{I}^{ij}}{m_0}\,,\qquad\beta_\text{P}^{ij}=\frac{\delta
  m_\text{P}^{ij}}{m_0}\,.\end{equation}
Obviously these parameters depend on the system under consideration.

By varying the Lagrangian~\eqref{L} we obtain the equation of motion
of the system as
\begin{equation}\label{EOM} \frac{\ud V^i}{\ud t} = \left(\delta^{ij} -
\beta_\text{I}^{ij}\right) \frac{\partial U}{\partial X^j} +
\beta_\text{P}^{jk} \frac{\partial U^{jk}}{\partial
  X^i}\,.\end{equation}
The system does not obey the weak equivalence principle (WEP) and the
parameters~\eqref{beta} can also be seen as WEP violating parameters,
respectively modifying the inertial (I) and passive (P) gravitational
masses of the system. Of course this means that the three different
aspects of the equivalence principle (WEP, LLI and LPI) are entangled
together by Schiff's conjecture~\cite{Schiff}. Finally the energy and
linear momentum of the system in the Newtonian approximation read
\begin{eqnarray} E &=& m_0 c^2 + \frac{1}{2} m_0\left(\delta^{ij} +
\beta_\text{I}^{ij}\right) V^i V^j - m_0\left(\delta^{ij} +
\beta_\text{P}^{ij}\right) \,U^{ij}(\bm{X})\,,\label{E}\\ P^i &=&
m_0\left(\delta^{ij} + \beta_\text{I}^{ij}\right)
V^j\,.\label{Pi}\end{eqnarray}

In the following we shall apply the modified Lagrangian~\eqref{L} to
the case of an atomic clock moving in the gravitational field of the
Sun or the Moon. The clock will be modelled by a two-level atom, and
of course we shall assume that the energy responsible for
the transition between levels in the atom is the abnormal internal
energy $E_\text{X}$.

\section{A simple Gedanken experiment}
\label{sect:simple}

\begin{figure}[t]
\begin{center}
\includegraphics[width=10cm]{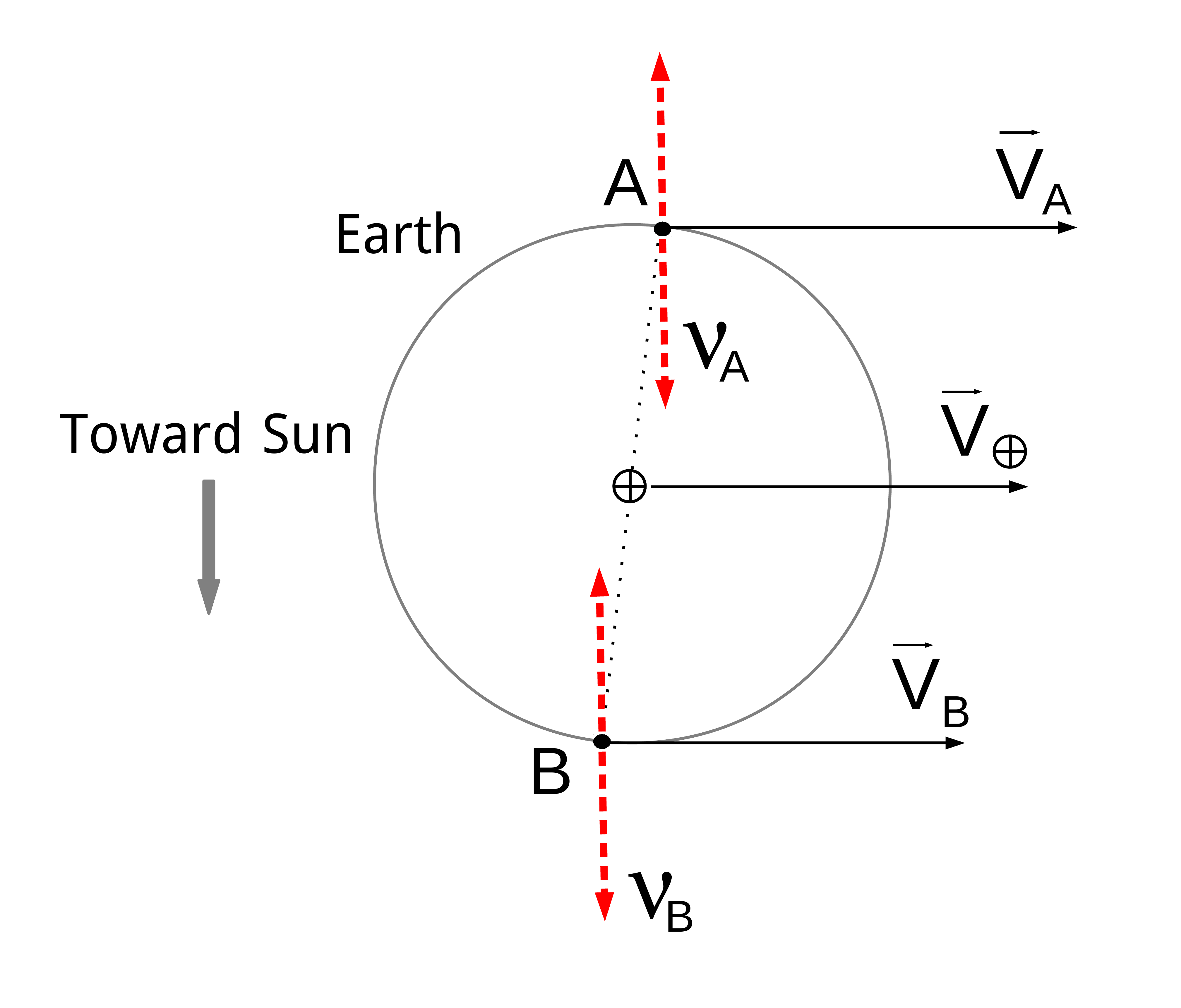}
\end{center}
\caption{Simplified experimental principle. Two photons $\nu_A$ are
  emitted from clock $A$ and compared to photons $\nu_B$ emitted from
  clock $B$. The velocities of $A$ and $B$ are perpendicular to the
  directions of emission of the photons. All positions and velocities
  are given in the Sun centred frame.}\label{fig1}
\end{figure}

In this section we examine a highly idealized situation in order to
simplify the calculations as much as possible, whilst allowing the
derivation of our main result. A full and general analysis is
presented in Sec.~\ref{sect:full}. We consider a perfectly spherical
Earth in circular orbit around a perfectly spherical Sun, with two
identical clocks fixed to the Earth's surface. We work
throughout in a Sun-centred, non rotating coordinate system. The
clocks are two level atoms, and a typical redshift experiment is
modelled as follows (see Fig.~\ref{fig1}): Clock $A$ undergoes a
transition from its excited state $\vert e\rangle$ to its ground state
$\vert g\rangle$ and emits a photon toward clock $B$. As the photon
passes $B$, clock $B$ undergoes a transition and emits a photon in the
same direction as the one incident from $A$. An observer then measures
the frequency difference between the two photons
(\textit{cf.}~\cite{Haugan}).
For simplicity in this section, we assume that the photon takes a
direct path from $A$ to $B$ through the Earth, that the atomic
transitions are two-photon transitions\footnote{Two-photon transitions
  involve initial and final atomic states whose difference in angular
  momentum is such that its conservation requires the
  absorption/emission of two photons \cite{Bass}.} with the two
photons emitted in opposite directions, and that the experiment is
arranged in such a way that the velocities of $A$ and $B$ (in the
Sun-centred frame) are perpendicular to the directions of the
respective emitted photons (see Fig.~\ref{fig1}).

Furthermore we will use a minimal form of the formalism described in
Sec.~\ref{sect:theory}, with $\delta m_\text{I}^{ij}=0$ and $\delta
m_\text{P}^{ij}=\delta m_\text{P}\delta^{ij}$, so the
Lagrangian~\eqref{L} is simplified to
\begin{equation}\label{L-simple} L = - m_0 c^2 + \frac{1}{2} m_0 
V^2 + \left(m_0+\delta m_\text{P}\right)U\,,
  \end{equation}
where $U=U(\bm{X})$ is the usual scalar Newtonian gravitational
potential.

\subsection{The photon frequency}
\label{sect:photon-simple}

We calculate the coordinate frequency $\nu$ of an emitted photon from
energy and momentum conservation between the initial and final
states. We write, to lowest order, the initial and final energies and
momenta of the system (atom $+$ photons) in the Sun centred frame,
using~\eqref{L-simple} and the atom's conserved
quantities~\eqref{E}--\eqref{Pi},
\begin{eqnarray}\label{E-p-LPI}
 E_f &=& m_0^g c^2+\frac{1}{2}m_0^g V_f^2-\bigl(m_0^g+\delta
 m^g_\text{P}\bigr) \,U + 2 h \nu\,, \\ E_i &=& m_0^e
 c^2+\frac{1}{2}m_0^e V_i^2-(m_0^e+\delta m_P^e) \,U\,,
\end{eqnarray}
where $h$ is Planck's constant, together with
\begin{equation}
P_f = m_0^g \,V_f\,, \qquad P_i = m_0^e \,V_i\,.
\end{equation}
We recall that $m_0^e=\overline{m}+\overline{E}_e/c^2$ and
$m_0^g=\overline{m}+\overline{E}_g/c^2$, with $\delta m_P^e$ and
$\delta m_P^g$ the abnormally coupled energy contributions to the
excited and ground states.

Momentum conservation yields to leading order $V_f=(1+\frac{\Delta
  \overline{E}}{\overline{m} c^2})V_i$ with $\Delta
\overline{E}=\overline{E}_e-\overline{E}_g$. Note that because of the
two-photon process the direction of $V$ is unchanged. To leading order
energy conservation then leads to
\begin{equation} \label{freq}
h \nu = h \nu_0\left(1-\frac{V_i^2}{2c^2} -
\frac{U}{c^2}\left(1+\alpha_\text{P}\right)\right)\,,
\end{equation}
where we pose $2h \nu_0=\Delta \overline{E}$ for the two-photon
  process, $\alpha_\text{P} = c^2 \Delta\delta m_\text{P}/\Delta
\overline{E}$ and $\Delta\delta m_\text{P}=\delta m^e_\text{P}-\delta
m^g_\text{P}$.

It is interesting to note that the initial and final velocities of the
atom are not equal. Indeed, this is required by momentum conservation
as the initial and final masses are unequal. The change of velocity is
directly related to aberration, \textit{i.e.}  although the photon
directions are perpendicular to the atom's velocity in the Sun centred
frame, they are not perpendicular in a frame at constant velocity
$V_i$, thus in that frame the atom receives a recoil ``kick'', with a
corresponding change of velocity in the Sun centred frame (of course,
the additional velocity is quickly dissipated in the clock
structure). Alternatively, one can arrange the experiment such that
the photons are emitted at a slight angle (\textit{i.e.} $\theta\sim
V_i/c$ in radians) in the Sun centred frame corresponding to
perpendicular emission in the $V_i$ frame. In that case the atom
velocity is unchanged. The corresponding calculation leads to the same
result as \eqref{freq}.

\subsection{Derivation of the redshift}

We calculate the coordinate frequency difference $(\nu_A-\nu_B)/\nu_0$
of the two co-located photons emitted at $A$ and $B$, respectively
(\textit{cf.} Fig.~\ref{fig1}). Applying Eq.~\eqref{freq} to the
emitted photons by $A$ and $B$ gives directly
\begin{equation} \label{freq-diff}
\frac{\nu_A-\nu_B}{\nu_0} =
\frac{U_B-U_A}{c^2}\left(1+\alpha_\text{P}\right) +
\frac{V_B^2-V_A^2}{2c^2}\,.
\end{equation}

For simplicity, we will assume that the Earth gravitational potential
is the same at $A$ and $B$ and we thus neglect it in the difference
$U_B-U_A$, and we also neglect small correction terms in $V_B^2-V_A^2$
due to the rotation of the Earth (see the next section for a fuller
treatment). The angular frequency of the Earth in circular orbit is
$\Omega^2 = (1+\beta_\text{P})GM_\odot/R_\oplus^{3}$, with $M_\odot$
the (reduced) mass of the Sun, $R_\oplus = \vert\bm{X}_\oplus\vert$
the Earth-Sun distance, and $\beta_\text{P}=\delta m_\text{P}/m_0$ for
the WEP violating parameter in the Sun field. The potentials are
$U_{A,B}=\frac{GM_\odot}{R_{A,B}}$ and the velocities $V_{A,B}^2 =
(\Omega R_{A,B})^2$. Then the frequency difference~\eqref{freq-diff}
becomes
\begin{equation}\label{freq-diff2}
\hspace{-1.5cm}\frac{\nu_A-\nu_B}{\nu_0} =
\frac{GM_\odot}{c^2}\left(\frac{1}{R_B} -
\frac{1}{R_A}\right)\left(1+\alpha_\text{P}\right) +
\frac{GM_\odot}{2c^2R_\oplus^3}\left(R_B^2 -
R_A^2\right)\left(1+\beta_\text{P}\right)\,.
\end{equation}
Writing $R_{A,B}=R_\oplus+\Delta R_{A,B}$ and expanding in terms of
the small quantities $\Delta R_{A,B}/R_\oplus$ we finally obtain
\begin{equation}\label{final}
\hspace{-1.5cm}\frac{\nu_A-\nu_B}{\nu_0} = \frac{GM_\odot}{c^2}\left(\frac{1}{R_B} -
\frac{1}{R_A}\right)\alpha_\text{P} +
\mathcal{O}\left(\frac{GM_\odot}{R_\oplus c^2}\frac{\Delta
  R_{A,B}^2}{R_\oplus^2}\right)\,,
\end{equation}
where we have neglected terms which are suppressed by a factor
$\beta_\text{P}/\alpha_\text{P} \lesssim \Delta
  \overline{E}/(m_\text{at}c^2)\ll 1$, with $m_\text{at}$ the average
  atomic mass of the elements making up the Earth [see
    Eq.~\eqref{alphabeta}]. When the equivalence principle is
satisfied ($\alpha_\text{P}=0$) one recovers the expected
result~\cite{H61,AW13} that the frequency difference is zero, up to
tidal terms. However, in presence of a violation of the equivalence
principle a frequency difference remains \textit{at first order in}
$\Delta U/c^2$, which is therefore testable experimentally.

\section{Full analysis of a realistic experiment}
\label{sect:full}

\subsection{The photon frequency}
\label{sect:photon}

We now work out the general case of the full modified
Lagrangian~\eqref{L} without simplifying assumptions regarding the
direction of propagation of the photons. We thus consider again the
transition between an initial excited $\vert e\rangle$ state toward a
final ground $\vert g\rangle$ state of the atom, accompanied by the
emission of a photon in the unit direction $\bm{N}=(N^i)$ as measured
in the global (Sun centred) frame. In particular, we now consider
  a one-photon process. Conservation of energy and linear momentum
between the initial and final states implies that
\begin{equation} \Delta E = h \nu\,,\qquad\Delta
\bm{P} = \frac{h \nu}{c}\bm{N}\,,\label{DeltaEP}
\end{equation}
where $\nu$ denotes the coordinate frequency of the emitted photon,
and $\Delta E=E_e-E_g$ and $\Delta \bm{P}=\bm{P}_e-\bm{P}_g$ are the
changes in the energy and linear momentum~\eqref{E}--\eqref{Pi}. Then
a simple calculation extending the one of Sec.~\ref{sect:simple},
combining both the energy and linear momentum conservation laws, thus
taking into account the recoil velocity $\Delta
\bm{V}=\bm{V}_e-\bm{V}_g$ of the atom during the transition, gives the
frequency of the photon as
\begin{equation}\label{hnu} h \nu = \frac{h \nu_0}{1 -
  \bm{N}\cdot\bm{V}/c}\left[ 1 - \frac{1}{2}\left(\delta^{ij} +
    \alpha_\text{I}^{ij} \right)\frac{V^i V^j}{c^2}- \left(\delta^{ij}
    + \alpha_\text{P}^{ij}
    \right)\frac{U^{ij}(\bm{X})}{c^2}\right]\,,\end{equation}
where we pose $h \nu_0=\Delta \overline{E}$ with $\Delta
\overline{E}=\overline{E}_e-\overline{E}_g$ and we recall that
$m_0^e=\overline{m}+\overline{E}_e/c^2$ and
$m_0^g=\overline{m}+\overline{E}_g/c^2$. The violation of the
universal redshift is here parametrized by the two parameters
\begin{equation}\label{alpha} 
\alpha_\text{I}^{ij} = \frac{\Delta \delta m_\text{I}^{ij}}{\Delta
  \overline{E}/c^2}\,,\qquad \alpha_\text{P}^{ij} = \frac{\Delta
  \delta m_\text{P}^{ij}}{\Delta \overline{E}/c^2}\,.\end{equation}
In addition we can compute the recoil velocity $\Delta \bm{V}$ as
\begin{equation}\label{recoilV} 
m_0 \Delta V^i = h \nu \left(\delta^{ij} - \beta_\text{I}^{ij}
\right)\frac{N^j}{c} - h \nu_0 \left(\delta^{ij} +
\alpha_\text{I}^{ij} - \beta_\text{I}^{ij}\right)
\frac{V^j}{c^2}\,.\end{equation}

These results are completely general, valid for a photon emitted in
any direction $\bm{N}$, and take in particular into account the recoil
of the atom. Notice that the recoil effect has two consequences on the
formula~\eqref{hnu}: (i) It implies the usual first-order Doppler
effect in the factor in front of the formula~\eqref{hnu}, and (ii) it
yields the correct sign for the second-order Doppler (or kinetic)
term. Even in standard GR (\textit{i.e.} independently from any EP
violation), taking properly into account the atom's recoil is crucial
in order to obtain the correct sign for the kinetic term in
Eq.~\eqref{hnu}. In that respect the corresponding formula in
Ref.~\cite{Haugan}, see Eq.~(2.22) there, and the formula (2.38) which
was replicated in Ref.~\cite{Will}, are incorrect as they do not
include the recoil of the atom, yielding the wrong sign for the
second-order Doppler effect.

The result~\eqref{hnu} gives us the coordinate frequency of the
emitted photon depending on the internal physics of the atom, and in
particular on the redshift violating parameters~\eqref{alpha}
associated with possible LLI and LPI violations of some internal
energy $E_\text{X}$ in the atom. It is well known~\cite{Nordtvedt}
that the WEP violating parameters~\eqref{beta} are related to their
redshift violating counterparts~\eqref{alpha} by
\begin{equation}\label{alphabeta}
\beta_\text{I}^{ij} \simeq \alpha_\text{I}^{ij}\frac{\overline{E}}{m_0
  c^2}\,,\qquad\beta_\text{P}^{ij} \simeq
\alpha_\text{P}^{ij}\frac{\overline{E}}{m_0 c^2}\,,
\end{equation}
so that WEP and redshift (or LPI/LLI) tests are not independent from
each other, but their relative interests depend on the details of the
model for the EP violation, and in particular on the type of abnormal
energy $E_\text{X}$ involved and the type of atom considered. The
  same conclusion arises also in different formalisms, such as the
  powerful Standard Model Extension (SME)~\cite{KS89,BK06,KT11}, and
  in formalisms motivated by string theory involving dilatonic or
  moduli scalar fields with gravitational strength, whose couplings to
  matter violate the EEP~\cite{Dequiv,DDono10}. (See also further
  comments in Sec.~\ref{sect:conclusion}.)

\subsection{Transformation to an Earth centred frame}
\label{sect:earth-frame}

The modified Lagrangian~\eqref{L} is defined in the global coordinate
system $(T, \bm{X})$ associated with the Sun --- centred on the Solar
System barycenter. In this section we shall need to apply a coordinate
transformation to a local coordinate system $(t, \bm{x})$ attached to
the Earth, and centred on the center of mass of the Earth.

Let $\bm{X}_\oplus(T)$ be the trajectory of the Earth around the Sun
in global coordinates. At a particular instant $T_0$ the position,
velocity and acceleration of the Earth are $\bm{X}_\oplus\equiv
\bm{X}_\oplus(T_0)$, $\bm{V}_\oplus\equiv\bm{V}_\oplus(T_0)$ and
$\bm{A}_\oplus\equiv\bm{A}_\oplus(T_0)$. The acceleration of the Earth
at that instant is $\bm{A}_\oplus = \bm{\nabla}U(\bm{X}_\oplus)$,
where $U$ is the gravitational potential of the Sun and/or other
bodies of the Solar System.\footnote{Note that the acceleration of the
  Earth should include terms in $\beta_\text{I}$ and $\beta_\text{P}$
  as in~\eqref{EOM}. However, in the final result these terms are
  suppressed by a factor $\beta/\alpha \ll 1$
  [\textit{cf.}~\eqref{alphabeta}], and can thus be neglected.}  In a
neighbourhood of the particular event $(T_0,\bm{X}_\oplus)$, we
\textit{define} the local (accelerated) coordinate system $(t,
\bm{x})$ centred on the Earth by
\begin{eqnarray} t &=& T - T_0 -
\frac{\bm{V}_\oplus\cdot\left(\bm{X} -
  \bm{X}_\oplus\right)}{c^2}\,,\label{tT}\\ \bm{x} &=& \bm{X} -
\bm{X}_\oplus - \bm{V}_\oplus\left(T-T_0\right) - \frac{1}{2}
\bm{A}_\oplus\left(T-T_0\right)^2 \,.\label{xX}
\end{eqnarray}
For the present analysis, essentially confined to Newtonian order,
such coordinate transformation will be sufficient. In particular we do
not need to consider the well-known acceleration term in the temporal
transformation law which would read
$t=(T-T_0)[1-(\bm{X}-\bm{X}_\oplus)\cdot\bm{A}_\oplus/c^2+\cdots]$, as
well as other terms coming from the Lorentz transformation. Only in
Sec.~\ref{sect:real}, where we consider an arbitrary satellite
velocity, will a non-Galilean term, the ${\cal O}(c^{-2})$ term
in~\eqref{tT}, be required. This term ensures that the speed of light
is an invariant within the required approximation
(see~\ref{sect:appendix}), and thus allows a correct calculation of
the first order Doppler effect in the local frame. Note in
  this respect that we are assuming that the violation of the
  equivalence principle affects the internal physics of the atom
  through the modified Lagrangian~\eqref{L}, but that the law of
  propagation of photons is ``standard''.

\begin{figure}[t]
\begin{center}
\includegraphics[width=12cm]{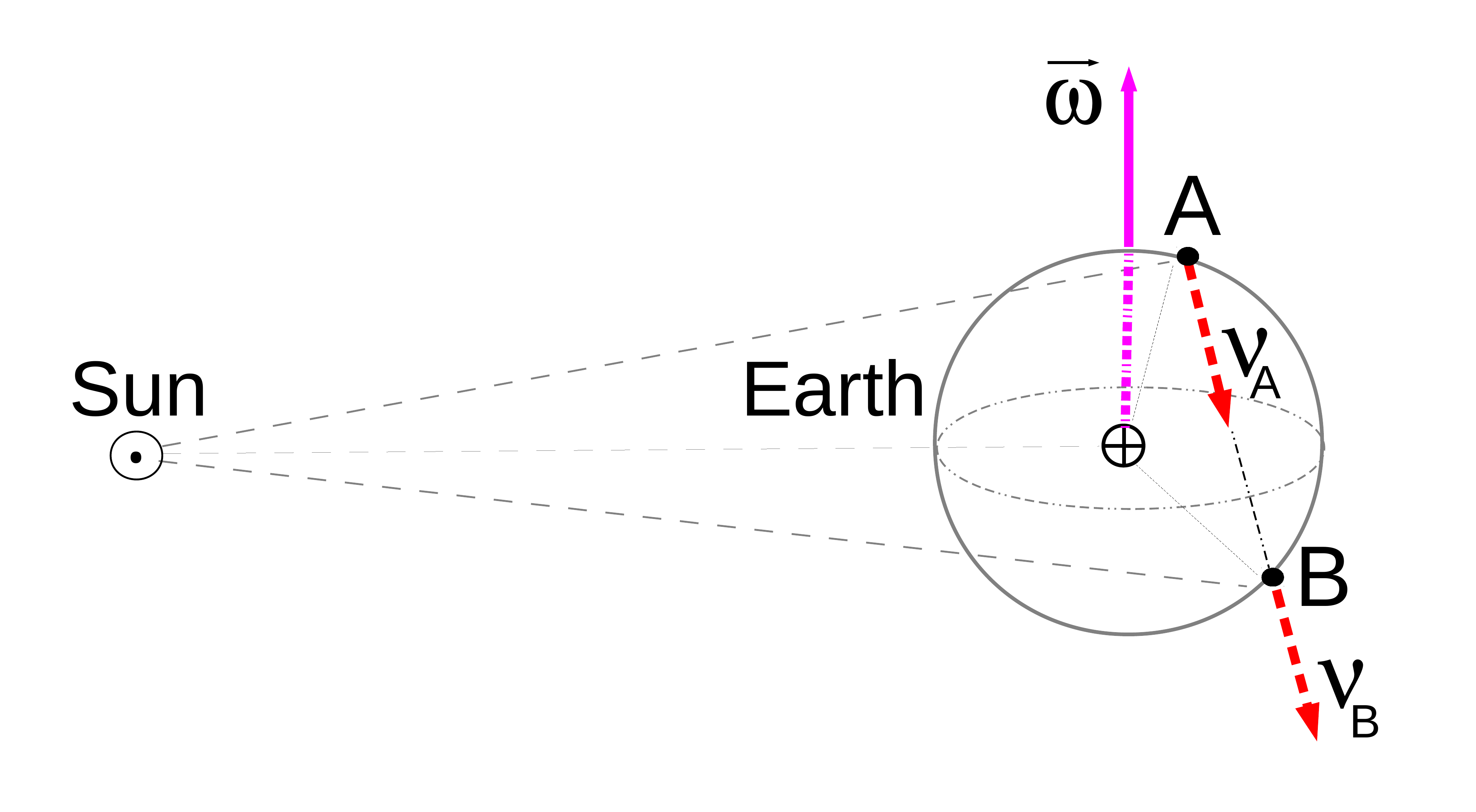}
\end{center}
\caption{The photon $\nu_{A}$ is emitted from ground clock $A$ toward
  $B$ and compared to the photon emitted from ground clock $B$ in the
  same direction as photon $\nu_{A}$.}\label{fig2}
\end{figure}
Consider the emission of a photon by an atomic transition at the
position $\bm{X}_A$ and instant $T_A$ in the global frame. This photon
propagates from $A$ to $B$ and is received at the position $\bm{X}_B$
and instant $T_B$. Simultaneously with the reception of this photon at
$T_B$, another photon is emitted by the same atomic transition of
another atom of the same species at point $\bm{X}_B$ in the same
direction as the photon coming from $A$ (see Fig.~\ref{fig2}). We want
to compute the difference of coordinate frequencies $\nu_A$ and
$\nu_B$ of these photons, obtained from Eq.~\eqref{hnu}.

For convenience we shall choose the instant $T_0$ to be the average
between the emission and reception instants, namely
\begin{equation}\label{T0} 
T_0 = \frac{T_A + T_B}{2}\,.\end{equation}
Defining also $\bm{R}_{AB} = \bm{X}_B(T_B) -\bm{X}_A(T_A)$ and the
associated unit direction $\bm{N}_{AB} = \bm{R}_{AB}/R_{AB}$ where
$R_{AB}=\vert\bm{R}_{AB}\vert$, we have $R_{AB}=c(T_B-T_A)$ for the
light-like separation. In the local frame the emission and reception
events $(t_A, \bm{x}_A)$ and $(t_B, \bm{x}_B)$ are given
by~\eqref{tT}--\eqref{xX}. In particular, with our choice of
``central'' instant~\eqref{T0}, the emission and reception in the
local frame occur at
\begin{eqnarray} t_A &=& - \frac{R_{AB}}{2 c} +
\calO\left(\frac{1}{c^2}\right)\,,\label{tA}\\ t_B &=& \frac{R_{AB}}{2
  c} +
\calO\left(\frac{1}{c^2}\right)\,,\label{tB}\end{eqnarray}
that are small quantities. At this stage we do not need to consider
the $1/c^2$ correction term in~\eqref{tT}. Using this, the spatial
positions in the local frame are then deduced
from~\eqref{tT}--\eqref{xX}. Neglecting terms $\sim 1/c^2$ we get
\begin{eqnarray} \bm{x}_A &=& \bm{X}_A - \bm{X}_\oplus +
\frac{R_{AB}}{2c} \bm{V}_\oplus +
\calO\left(\frac{1}{c^2}\right)\,,\\ \bm{x}_B &=& \bm{X}_B -
\bm{X}_\oplus - \frac{R_{AB}}{2c} \bm{V}_\oplus +
\calO\left(\frac{1}{c^2}\right)\,.
\end{eqnarray}
By differentiating~\eqref{tT}--\eqref{xX} we compute the corresponding
velocities as
\begin{eqnarray}
\bm{v}_A &=& \bm{V}_A - \bm{V}_\oplus + \frac{R_{AB}}{2c}
\bm{A}_\oplus + \calO\left(\frac{1}{c^2}\right)\,,\\ \bm{v}_B &=&
\bm{V}_B - \bm{V}_\oplus - \frac{R_{AB}}{2c} \bm{A}_\oplus +
\calO\left(\frac{1}{c^2}\right)\,.
\end{eqnarray}

The main point of our calculation is to implement the fact that the
two emitting and receiving atoms/clocks are attached to the
Earth. This assumption will be implemented in the most general way by
imposing that \textit{in the local frame} $(t, \bm{x})$ defined by
Eqs.~\eqref{tT}--\eqref{xX} the motion of the atom/clock is that of a
rigid rotator, \textit{i.e.}
\begin{equation}\label{rigid}
\bm{v}_A = \bm{\omega} \times \bm{x}_A \,,\qquad\bm{v}_B = \bm{\omega}
\times \bm{x}_B \,,
\end{equation}
where $\bm{\omega}$ denotes the rotation vector of the Earth supposed
to be constant. Here, consistent with our Newtonian approximation, we
only need to consider the usual rigid rotation
conditions~\eqref{rigid}.\footnote{See \textit{e.g.}~\cite{Bona83} and
  references therein for discussions on rigidity conditions
  in special relativity.}  Translated into the global frame $(T,
\bm{X})$ the latter assumption imposes that
\begin{eqnarray} \bm{V}_A - \bm{V}_\oplus &=& \bm{\omega}
\times \left(\bm{X}_A - \bm{X}_\oplus\right) + \frac{R_{AB}}{2c}
\Bigl[ \bm{\omega} \times \bm{V}_\oplus - \bm{A}_\oplus \Bigr] +
\calO\left(\frac{1}{c^2}\right)\,,\\ \bm{V}_B - \bm{V}_\oplus &=&
\bm{\omega} \times \left(\bm{X}_B - \bm{X}_\oplus\right) -
\frac{R_{AB}}{2c} \Bigl[ \bm{\omega} \times \bm{V}_\oplus -
  \bm{A}_\oplus \Bigr] + \calO\left(\frac{1}{c^2}\right)\,.
\end{eqnarray}
Substracting those equations we get for the relative velocity,
\begin{equation}\label{rel}
\bm{V}_B - \bm{V}_A = \bm{\omega} \times \bm{R}_{AB} -
\frac{R_{AB}}{c} \Bigl[ \bm{\omega} \times \bm{V}_\oplus -
  \bm{A}_\oplus \Bigl] + \calO\left(\frac{1}{c^2}\right)\,.
\end{equation}

\subsection{Derivation of the redshift}
\label{sect:tidal}

We now use the previous relations and the result~\eqref{hnu} to
explicitly compute the frequency shift between the photons $A$ and
$B$. Let us first recover the standard GR prediction, setting
$\alpha_\text{I}^{ij}$ and $\alpha_\text{P}^{ij}$ in~\eqref{hnu} to
zero. Expanding to the required order we obtain
\begin{eqnarray}\label{expand} 
&&\hspace{-1.5cm}\left(\frac{\nu_A - \nu_B}{\nu_0}\right)_{\rm{GR}} =
  \frac{1}{c}\bm{N}_{AB}\cdot\left(\bm{V}_A-\bm{V}_B\right)
  \nonumber\\&&\hspace{-1.0cm}
  +\frac{1}{c^2}\left[(\bm{N}_{AB}\cdot\bm{V}_A)^2
    -(\bm{N}_{AB}\cdot\bm{V}_B)^2-\frac{\bm{V}_A^2}{2}-U_A
    +\frac{\bm{V}_B^2}{2}+U_B\right] +
  \calO\left(\frac{1}{c^3}\right)\,.
\end{eqnarray}
Now, thanks to~\eqref{rel}, we see that the first term in that
expression, which is the usual first order Doppler effect, is actually
of second order. We then obtain, in a first stage,\footnote{Note that
  $(\bm{N}_{AB}\cdot\bm{V}_A)^2-(\bm{N}_{AB}\cdot\bm{V}_B)^2 =
  [\bm{N}_{AB}\cdot(\bm{V}_B - \bm{V}_A)][\bm{N}_{AB}\cdot(\bm{V}_B +
    \bm{V}_A)]=\calO\left(\frac{1}{c}\right)$ from~\eqref{rel}.}
\begin{equation}\label{redshift0} 
\hspace{-2.5cm}\left(\frac{\nu_A - \nu_B}{\nu_0}\right)_{\rm{GR}} =
\frac{1}{c^2}\left[U_B - U_A +
  \frac{\bm{V}_B^2}{2}-\frac{\bm{V}_A^2}{2} + \bm{R}_{AB}\cdot\bigl(
  \bm{\omega} \times \bm{V}_\oplus - \bm{A}_\oplus \bigr)\right] +
\calO\left(\frac{1}{c^3}\right)\,.
\end{equation}
In the small $1/c^2$ terms we can approximate $\bm{X}_A=\bm{X}_\oplus
+ \bm{x}_A$ and $\bm{X}_B=\bm{X}_\oplus + \bm{x}_B$, and expand the
Newtonian potentials $U_A$ and $U_B$ around the value at the center of
the Earth, $U_\oplus=U(\bm{X}_\oplus)$, with higher order terms of
that expansion being the tidal terms. For simplicity we keep only the
dominant tidal term at the quadrupole level. Finally the net
prediction from GR (and the Einstein equivalence principle) reads
as\footnote{Intermediate formulas useful in this calculation are
\begin{eqnarray*}
U_B - U_A &=&
\frac{1}{2}\left(x_B^ix_B^j-x_A^ix_A^j\right)\frac{\partial^2U}{\partial
  x^i\partial x^j}(\bm{X}_\oplus) +
\bm{R}_{AB}\cdot\bm{A}_\oplus\,,\\\frac{\bm{V}_B^2}{2}-\frac{\bm{V}_A^2}{2}
&=& \frac{\bm{v}_B^2}{2}-\frac{\bm{v}_A^2}{2} +
\left(\bm{\omega}\times\bm{R}_{AB}\right)\cdot\bm{V}_\oplus\,.
\end{eqnarray*}}
\begin{equation}\label{redshiftGR}
\hspace{-1.5cm}\left(\frac{\nu_A - \nu_B}{\nu_0}\right)_{\rm{GR}} =
\frac{1}{2c^2}\Biggl[
  \left(x_B^ix_B^j-x_A^ix_A^j\right)\frac{\partial^2U}{\partial
    x^i\partial x^j}(\bm{X}_\oplus) + \bm{v}_B^2-\bm{v}_A^2\Biggr] +
\calO\left(\frac{1}{c^3}\right)\,.
\end{equation}
Reminiscent of the absence of Noon-Midnight redshift, the latter GR
effect is very small, as it scales like $\sim G M_\odot
r_\text{E}^2/(R_\oplus^3 c^2)$ and is typically of the order of
$10^{-17}$. Evidently this is because the Earth is freely falling in
the field of the Sun (and of other bodies of the Solar System), so the
redshift depends only on the tidal field of the Sun rather than on the
field itself. In the freely falling frame the laws of special
relativity hold and there is no redshift between clocks. Furthermore
the rigid rotation of the Earth does not give rise to an effect
either. This can easily be checked in the simple configuration
analyzed in Sec.~\ref{sect:simple}. This is due to the fact that
photons do not experience any redshift when their emitters and
receivers are attached to the rim of a centrifuge in special
relativity (see a classic exercise on p.~63 of MTW~\cite{MTW}).

Finally we complete our analysis by simply adding the contributions of
the redshift violating parameters $\alpha_\text{I}^{ij}$ and
$\alpha_\text{P}^{ij}$ that are immediately seen from Eq.~\eqref{hnu}
to result in
\begin{equation}\label{finalredshift} 
\hspace{-1.5cm}\frac{\nu_A - \nu_B}{\nu_0} = \left(\frac{\nu_A -
  \nu_B}{\nu_0}\right)_{\rm{GR}}
+\frac{1}{c^2}\Biggl[\alpha_\text{I}^{ij}\frac{V_B^iV_B^j-V_A^iV_A^j}{2}
  + \alpha_\text{P}^{ij}\left(U_B^{ij}-U_A^{ij}\right)\Biggr]\,.
\end{equation}
The non-GR terms depend on the velovities $V^i_{A,B}$ in the global
frame, since the EP violating parameters $\alpha_\text{I}^{ij}$ and
$\alpha_\text{P}^{ij}$ are defined in that frame,
see~\eqref{hnu}. Note that in the left-hand side
of~\eqref{finalredshift} the coordinate frequencies $\nu_{A,B}$ should
be the ones measured in the local frame. At that order, because the
final effect~\eqref{finalredshift} is already of order $\sim 1/c^2$,
we do not need to correct for the coordinate frequencies taking into
account the $\mathcal{O}(c^{-2})$ term in the transformation
law~\eqref{tT}. However, in Sec.~\ref{sect:real}, where we investigate
a more realistic case of the comparison of the two clocks $A$ and $B$
\textit{via} a satellite $S$, we shall need to include that term to
transform the frequencies $\nu_{A,B}$.

The same formula applies to any external body in the Solar System, but
with the distinction that in the case of a violation of the EP the
redshift parameter $\alpha_\text{P}^{ij}$ is expected to depend on the
particular source of the gravitational field (Sun, Moon,
\textit{etc.}). Thus its contribution to Eq.~\eqref{finalredshift}
should rather be a sum over all possible bodies $n$ with different EP
violating parameters $(\alpha_\text{P}^{ij})_n$:
\begin{equation}\label{finalredshift2} 
\hspace{-1.5cm}\frac{\nu_A - \nu_B}{\nu_0} = \left(\frac{\nu_A -
  \nu_B}{\nu_0}\right)_{\rm{GR}}
+\frac{1}{c^2}\Biggl[\alpha_\text{I}^{ij}\frac{V_B^iV_B^j-V_A^iV_A^j}{2}
  + \sum_n
  (\alpha_\text{P}^{ij})_n\left(U_B^{ij}-U_A^{ij}\right)_n\Biggr]\,.
\end{equation}

As we can see, despite the fact that the GR contribution is extremely
small due to the Earth freely falling toward the Sun, the non-GR LPI
violating corrections in~\eqref{finalredshift}--\eqref{finalredshift2}
are linear in the potentials of the exterior bodies. This fact allows
to test at an interesting level in Earth vicinity the redshift
parameters $\alpha_\text{I}^{ij}$ and $(\alpha_\text{P}^{ij})_n$. For
instance, one expects to obtain in the case of the STE-QUEST
experiment~\cite{STEQUEST} a test of the gravitational redshift due to
the Sun to an uncertainty of $\vert\alpha_\text{P}\vert \leqslant
2\times 10^{-6}$, with an ultimate goal of $5\times 10^{-7}$. For the
case of the Moon, the expected uncertainty should be $4\times
10^{-4}$, with an ultimate goal of $9\times 10^{-5}$. Since the
measurement will consist in the indirect comparison of signals from
two clocks located on the ground \textit{via} the satellite orbiting
the Earth, we present in the next subsection a more realistic analysis
with a satellite $S$ linked to the two ground clocks $A$ and $B$.

\subsection{Two ground clocks compared via a satellite}
\label{sect:real}

\begin{figure}[t]
\begin{center}
\includegraphics[width=13cm]{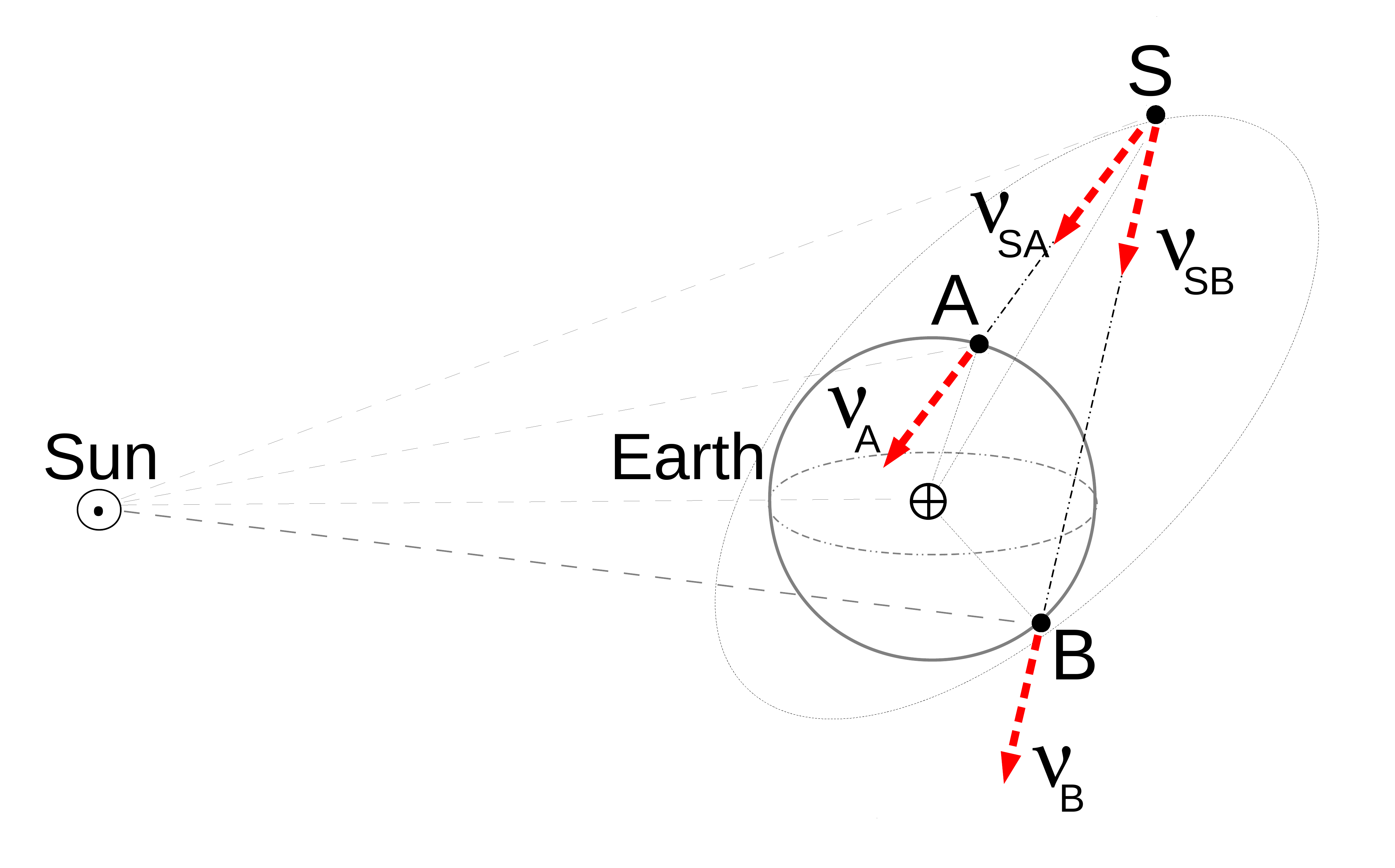}
\end{center}
\caption{Two photons $\nu_{SA}$ and $\nu_{SB}$ are emitted from a
  clock on board a satellite $S$ and compared to photons emitted in
  the same directions from ground clocks $A$ and $B$.}\label{fig3}
\end{figure}

The experimental configuration is shown in Fig.~\ref{fig3}. The
emitter is now on-board the satellite $S$, and the emitted photons
$\nu_{SA}$ and $\nu_{SB}$ are compared to those emitted by the two
ground clocks $A$ and $B$. The two photons are emitted from the
satellite at the same instant, thus all quantities related to the
satellite (velocity, gravitational potential, instrumental noise and
biases) are the same for both photons. This ``common view''
arrangement is analogous to the actual situation planned for
STE-QUEST, whose orbit is designed precisely to allow for long common
view periods between ground clocks located on different continents.

The corresponding frequency difference is calculated from the
individual links by
\begin{equation} \label{freq-S}
\frac{\nu_A - \nu_B}{\nu_0} = \frac{\nu_{SB} - \nu_B}{\nu_0}-
\frac{\nu_{SA} - \nu_A}{\nu_0}\,.
\end{equation}
The clock on board the satellite is not fixed to the Earth an moves at
arbitrary velocity, thus Eq.~\eqref{finalredshift2} cannot be directly
applied to the individual links as the assumption~\eqref{rigid} is not
satisfied for $S$. Instead, going back to~\eqref{hnu} and
concentrating first on the GR part, we get the same result as
in~\eqref{expand},
\begin{eqnarray}\label{expand-S} 
&&\hspace{-1.5cm}\left(\frac{\nu_{SA} - \nu_A}{\nu_0}\right)_{\rm{GR}} =
  \frac{1}{c}\bm{N}_{SA}\cdot\left(\bm{V}_S-\bm{V}_A\right)
  \nonumber\\&&\hspace{-1.0cm}
  +\frac{1}{c^2}\left[(\bm{N}_{SA}\cdot\bm{V}_S)^2
    -(\bm{N}_{SA}\cdot\bm{V}_A)^2-\frac{\bm{V}_S^2}{2}-U_S
    +\frac{\bm{V}_A^2}{2}+U_A\right] +
  \calO\left(\frac{1}{c^3}\right)\,,
\end{eqnarray}
with a similar expression for the comparison of $S$ to $B$. We will
aim at expressing all quantities in the local frame, in order to
recover the standard expressions, in particular for the first order
Doppler shifts. This requires converting the photon coordinate
frequencies from the global frame, as is used in Eq.~\eqref{expand-S},
to the local frame. As shown in~\ref{sect:appendix} the frequency
difference transforms as
\begin{eqnarray}\label{transfo}
\hspace{-1.5cm}\left(\frac{\nu_{SA} -
  \nu_A}{\nu_0}\right)^{\rm{local}}_{\rm{GR}} &=& \left(\frac{\nu_{SA}
  - \nu_A}{\nu_0}\right)_{\rm{GR}} \nonumber \\ &-&
\frac{1}{c^2}(\bm{n}_{SA}\cdot\bm{V}_\oplus)
\Bigl(\bm{N}_{SA}\cdot(\bm{V}_S-\bm{V}_A)\Bigr) +
\calO\left(\frac{1}{c^3}\right),
\end{eqnarray}
with a similar expression for the comparison of $S$ to $B$. Note that
with both clocks fixed on the Earth the correction term
in~\eqref{transfo} is of order ${\cal O}(c^{-3})$ thanks
to~\eqref{rel}, which justifies not having used~\eqref{transfo} in the
previous sections.

Substituting~\eqref{expand-S} and~\eqref{transfo} into~\eqref{freq-S},
the redshift and second-order Doppler terms due to the satellite
cancel leaving
\begin{eqnarray}\label{redshift-S} 
\hspace{-2.5cm}\left(\frac{\nu_{A} -
  \nu_B}{\nu_0}\right)^{\rm{local}}_{\rm{GR}} &=&
\frac{1}{c}\Bigl[\bm{N}_{SB}\cdot
  \left(\bm{V}_S-\bm{V}_B\right)-\bm{N}_{SA}\cdot
  \left(\bm{V}_S-\bm{V}_A\right)\Bigr] \nonumber\\ \hspace{-1.5cm}
&+&\frac{1}{c^2}\Biggl[(\bm{N}_{SB}\cdot\bm{V}_S)^2
  -(\bm{N}_{SB}\cdot\bm{V}_B)^2-(\bm{N}_{SA}\cdot\bm{V}_S)^2
  +(\bm{N}_{SA}\cdot\bm{V}_A)^2 \nonumber \\ && +
  (\bm{n}_{SA}\cdot\bm{V}_\oplus)
  \Bigl(\bm{N}_{SA}\cdot(\bm{V}_S-\bm{V}_A)\Bigr) -
  (\bm{n}_{SB}\cdot\bm{V}_\oplus)
  \Bigl(\bm{N}_{SB}\cdot(\bm{V}_S-\bm{V}_B)\Bigr) \nonumber \\&& +
  \frac{\bm{V}_B^2}{2}+U_B -\frac{\bm{V}_A^2}{2}-U_A
  \Biggr]+\calO\left(\frac{1}{c^3}\right)\,.
\end{eqnarray}
The first two lines result from (the expansion of) the first order
Doppler effect, the third line comes from the
transformation~\eqref{transfo} and the last line is identical to the
corresponding terms in~\eqref{expand}. We now
use~\eqref{tT}--\eqref{xX} to transform all $\bm{N}$'s and $\bm{V}$'s
to the Earth centred frame, with $T_0=T_S$ (the emission time on the
satellite) instead of~\eqref{T0}.\footnote{Note in particular that, as
  shown in \ref{sect:appendixB} (and similarly for $\bm{N}_{SB}$
  \textit{vs.} $\bm{n}_{SB}$)
$$\bm{N}_{SA} =
    \bm{n}_{SA}\Bigl(1-\frac{1}{c}\bm{n}_{SA}\cdot\bm{V}_\oplus\Bigr)
    + \frac{1}{c}\bm{V}_\oplus\,.$$} We also apply the
ansatz~\eqref{rigid} to the two ground clocks $A$ and $B$ (but not to
the satellite clock $S$). This leads after some vector algebra similar
to Sec.~\ref{sect:tidal} to a result analogous to~\eqref{redshiftGR},
\begin{equation}\label{redshiftGR-S}
\hspace{-1.5cm}\left(\frac{\nu_A -
  \nu_B}{\nu_0}\right)^{\rm{local}}_{\rm{GR}} = \Delta_S +
\frac{1}{2c^2}\Biggl[
  \left(x_B^ix_B^j-x_A^ix_A^j\right)\partial_{ij}U_\oplus +
  \bm{v}_B^2-\bm{v}_A^2\Biggr] + \calO\left(\frac{1}{c^3}\right)\,,
\end{equation}
with, however, some extra terms representing first-order Doppler
effects in the Earth centred frame, which are now non-zero, contrary
to Eq.~\eqref{redshiftGR}, as the satellite clock $S$ is not fixed on
the Earth surface. These extra terms are the usual (expansion of) the
first order Doppler effect given by
\begin{eqnarray}\label{Delta-S} 
\Delta_S &=& \frac{1}{c}\bigl[\bm{n}_{SB}\cdot\left(\bm{v}_S -
  \bm{v}_B\right)-\bm{n}_{SA}\cdot\left(\bm{v}_S-\bm{v}_A\right)\bigr]
\nonumber\\ &+&\frac{1}{c^2}\Bigl[(\bm{n}_{SB}\cdot\bm{v}_S)^2
  -(\bm{n}_{SB}\cdot\bm{v}_B)^2-(\bm{n}_{SA}\cdot\bm{v}_S)^2
  +(\bm{n}_{SA}\cdot\bm{v}_A)^2\Bigr]\,.
\end{eqnarray}
As expected Eq.~\eqref{redshiftGR-S} only includes tidal terms in the
gravitational redshift, which are small. However, the non-GR terms are
given exactly like in the previous section by~\eqref{finalredshift}
or~\eqref{finalredshift2} and can be used to bound non-Einsteinian EP
violating parameters at leading order in $U/c^2$ as previously
discussed.

\section{Conclusion}
\label{sect:conclusion}

We have shown that in a very broad class of equivalence principle
violating theories, a test of the gravitational redshift in the field
of external bodies (Sun, Moon, ...) using Earth fixed clocks is
sensitive {\it to first order} in $\Delta U_\text{ext}/c^2$ to a
possible violation. In doing so we have also provided a detailed
derivation of the well known result (sometimes coined the absence of
Noon-Midnight redshift) that in the GR limit only tidal terms in the
gravitational redshift can be observed, which are a factor $\sim
r_\text{E}/R_\oplus$ smaller. The calculations were carried out
consistently for different configurations, including the comparison of
ground clocks \textit{via} a satellite as planned in the STE-QUEST
mission.

We modelled the clocks by two-level atoms and derived the coordinate
frequency of the photon emitted by an atomic transition, depending on
the internal physics of the atom and in particular on some possible
abnormal dependence of the internal energy on the position and
velocity of the atom. In the modified Lagrangian
formalism~\cite{Dicke,Nordtvedt,Haugan,Will} (see
  also~\cite{KS89,BK06,KT11,Dequiv,DDono10} for alternative
  formalisms), such dependence is parametrized by EP violating
parameters associated with the violation of the local Lorentz
invariance (LLI) and the local position invariance (LPI). We pointed
out the importance of taking into account in this formalism the recoil
of the atom in order to recover the usual first-order Doppler effect
and the correct sign for the second-order Doppler correction.

In addition to the test of the gravitational redshift, the STE-QUEST
satellite will also carry an experiment measuring the weak equivalence
principle (WEP) or universality of free fall at the level $10^{-15}$
by means of dual-species atomic interferometry~\cite{STEQUEST}. Let us
thus finish by briefly discussing the relative merits of different
types of tests of the equivalence principle in our theoretical
framework. As mentioned in Sec.~\ref{sect:photon}, tests of the
gravitational redshift --- LPI aspect of the equivalence principle ---
as well as tests of LLI are related to tests of WEP \textit{via} a
factor $\overline{E}/(m_0 c^2)$,
\textit{cf.}~\eqref{alphabeta}. Therefore, the comparison between the
different tests depends crucially on the details of the chosen model
for the violation of the equivalence principle. For example, if all kinds of electromagnetic energy (including e.g. nuclear binding energy) contribute to the equivalence principle violation then
$\overline{E}/(m_0 c^2)\approx 10^{-3}$. But if only nuclear spin
plays a role then the involved energies are those of the hyperfine
transitions (GHz frequencies) and $\overline{E}/(m_0 c^2)\approx
10^{-16}$. Thus WEP and gravitational redshift tests compare
differently by many orders of magnitude depending on the detailed
model used. For instance, see Ref.~\cite{Dequiv} for a comparison
  between WEP and redshift/clock tests using generic dilatonic EP
  violating models. The most reasonable strategy in a general search
for the violation of the equivalence principle is to perform the test
of the gravitational redshift alongside with the test of the WEP. This
is the driving motivation of the STE-QUEST mission.

\ack Several discussions with Stefano Vitale and Clifford Will during
or after the assessment study phase of STE-QUEST prompted the analysis
performed in this paper. PW acknowledges helpful discussions with
Pac\^ome Delva and members of the ESA Physical Sciences Working Group.

\appendix

\section{Photon coordinate frequency in the local frame}
\label{sect:appendix}

We work in the geometric optics approximation with the photon's
four-vector $K_\mu=\partial\phi/\partial X^\mu =
  (-\Omega/c_\text{global},\bm{K})$ in the global frame $(T, \bm{X})$,
  where as usual $\Omega = 2\pi\nu_\text{global}$ and $\bm{K}=K\bm{N}$
  with $K=\vert\bm{K}\vert=\Omega/c_\text{global}$. Similarly in the
  local frame $(t, \bm{x})$ defined by Eqs.~\eqref{tT}--\eqref{xX}, we
  have $k_\mu=\partial\phi/\partial x^\mu =
  (-\omega/c_\text{local},\bm{k})$ where $\omega =
  2\pi\nu_\text{local}$ with $\bm{k}=k\bm{n}$ and
  $k=\omega/c_\text{local}$.

As already mentioned in Sec.~\ref{sect:earth-frame}, the extra term
$\calO(c^{-2})$ in the temporal transformation
  law~\eqref{tT}, where for definiteness we identify $c\equiv
  c_\text{global}$, ensures that the speed of light is invariant to
leading order, \textit{i.e.}
$c_\text{local}=c_\text{global}[1+\calO(c^{-2})]$ or, more
  precisely,
\begin{equation}\label{clocal}
c_\text{local}=c_\text{global}\!\left[1+{\cal
    O}\left(\frac{V^2_\oplus}{c^2},
  \frac{A_\oplus\vert\bm{X}-\bm{X}_\oplus\vert}{c^2}\right)\right]\,,
\end{equation}
which implies that in the local frame
$k=-k_0=\omega/c_\text{global}$ modulo small corrections
$\mathcal{O}(c^{-3})$. Using the transformations~\eqref{tT}
and~\eqref{xX} we easily find to first order in $V_\oplus/c$, $K_0 =
\frac{\partial x^\mu}{\partial X^0}k_\mu =
k_0-\bm{k}\cdot\bm{V}_\oplus/c$, or equivalently
\begin{equation}\label{nuNu}
\nu_\text{global} =
\nu_\text{local}\left[1+\frac{\bm{n}\cdot\bm{V_\oplus}}{c}+
  \calO\left(\frac{1}{c^2}\right)\right]\,.
\end{equation}
We now apply~\eqref{nuNu} to, for example, the satellite to ground
normalized frequency difference to obtain directly
\begin{equation}\label{dfDF}
\left(\frac{\nu_{SA}-\nu_A}{\nu_0}\right)_\text{global} =
\left(\frac{\nu_{SA}-\nu_A}{\nu_0}\right)_\text{local} \left[ 1 +
  \frac{\bm{n}_{SA}\cdot\bm{V_\oplus}}{c} +
  \calO\left(\frac{1}{c^2}\right)\right]\,.
\end{equation}
In the correction term we can replace $(\nu_{SA}-\nu_A)/\nu_0$ by the
leading term of~\eqref{expand-S} which leads directly
to~\eqref{transfo}.

\section{Transformation of unit direction vectors}
\label{sect:appendixB}

We use the transformation~\eqref{xX} with $T_0=T_S$ (the emission time
on the satellite) to transform the unit vector $\bm{N}_{SA}$. Working
to only first order in $V/c$, Eq.~\eqref{xX} leads to
\begin{equation}
\bm{X}_{S} = \bm{x}_{S}+\bm{X}_{\oplus}\,, \qquad\bm{X}_{A} =
\bm{x}_{A}+\bm{X}_{\oplus}+\frac{R_{SA}}{c}\bm{V}_{\oplus} +
\calO\left(\frac{1}{c^2}\right)\,,\label{XxAppendix}
\end{equation}
where we have used $T_{A}-T_{S} = R_{SA}/{c}$. From~\eqref{XxAppendix}
we obtain
\begin{equation}
\label{RSAappendix}
\hspace{-2cm}R_{SA}=|\bm{X}_{A}-\bm{X}_{S}|=|\bm{x}_{A}-\bm{x}_{S} +
\frac{R_{SA}}{c}\bm{V}_{\oplus}| =
r_{SA}\left(1+\frac{\bm{n}_{SA}\cdot\bm{V}_{\oplus}}{c}\right) +
\calO\left(\frac{1}{c^2}\right)\,,
\end{equation}
where $\bm{r}_{SA}=\bm{x}_{A}-\bm{x}_{S}$ and
$\bm{n}_{SA}=\bm{r}_{SA}/r_{SA}$. Using~\eqref{XxAppendix}
and~\eqref{RSAappendix} then directly leads to
\begin{equation}
\label{NSAappendix}
\bm{N}_{SA}=\frac{\bm{X}_{A}-\bm{X}_{S}}{R_{SA}} =
\bm{n}_{SA}\Bigl(1-\frac{\bm{n}_{SA}\cdot\bm{V}_{\oplus}}{c}\Bigr) +
\frac{1}{c}\bm{V}_\oplus+\calO\left(\frac{1}{c^2}\right)\,.
\end{equation}

\section*{References}

\bibliography{ListeRef}

\end{document}